\documentstyle[epsf]{l-aa}  
\begin{document}
\newcommand{\ang}{$\rm \AA$}
\newcommand{\zm}{M$_{\odot}$}
\newcommand{\zl}{L$_{\odot}$}
\newcommand{\arc}{$^{\prime\prime}$ }
\newcommand{\mic}{$\mu$m}
\newcommand{\ea}{et al.}
\newcommand{\hra}{HR~1865}
\newcommand{\hrz}{HR~1017}
\newcommand{\irv}{IRAS~04296+3429}
\newcommand{\irt}{IRAS~22223+4327}
\newcommand{\km}{$\rm{km\,s^{-1}}$}
\newcommand{\ha}{H$_{\alpha}$}
\newcommand{\apj}[1]{{ApJ }{ #1}}
\newcommand{\apjss}[1]{{ApJS }{ #1}}
\newcommand{\astronj}[1]{{AJ }{ #1}}
\newcommand{\aea}[1]{{A\&A }{ #1}}
\newcommand{\aeass}[1]{{A\&AS }{ #1}}
\newcommand{\mnras}[1]{{MNRAS }{ #1}}
\newcommand{\acas}[1]{{Acta Astron.\ }{ #1}}
\newcommand{\pasp}[1]{{PASP }{ #1}}
\newcommand{\pasj}[1]{{Publ.\ Astron.\ Soc.\ Japan\ }{ #1}}
\newcommand{\araa}[1]{{ARA\&A }{ #1}}
\newcommand{\iau}{1993 in ``IAU Symposium No. 155 Planetary Nebulae"
 eds. Weinberger R. and Acker A., Kluwer academic publishers, The
 Netherlands,}
\newcommand{\chili}{1993 in ``Second ESO/CTIO workshop Mass loss on the
 AGB and beyond" ed. Schwarz H.E., ESO Conference and Workshop Proceedings
No. 46, Germany,}
\newcommand{\harvard}{1993 in  ``Luminous high latitude stars", ed. Sasselov
 S.S., Astronomical society of the Pacific conference series Vol. 45,
ASP, USA,}

\thesaurus{08.09.2:HR~1017, HR~1865, IRAS~04296+3429, IRAS~22223+4327 -- 08.16.4
-- 08.01.1 }
\title{
Photospheric composition of the carbon-rich 21 \mic\ post-AGB stars
 IRAS~22223+4327 and IRAS~04296+3429
\thanks{Based on observations with the Utrecht Echelle Spectrograph, 
fed by the 4.2m William Herschel Telescope at La Palma}}

\author{Leen Decin\inst{1}\thanks{Scientific researcher of the Fund for
Scientific Research, Flanders}, Hans Van Winckel\inst{1}\thanks{Postdoctoral fellow 
of the Fund for Scientific 
Research, Flanders}, Christoffel Waelkens\inst{1}, Eric J. Bakker\inst{2}}

\offprints{Leen Decin. e-mail Leen.Decin@ster.kuleuven.ac.be}

\institute{Instituut voor Sterrenkunde, Katholieke Universiteit Leuven,
Celestijnenlaan 200B, B-3000 Leuven, Belgium \and Department of Astronomy
and McDonald Observatory, University of Texas, Austin TX78712, USA }

\date{received,  accepted}
\maketitle
\markboth{Leen Decin et al.: Optical spectroscopy of carbon-rich 
post-AGB stars}{Leen Decin et al.: Optical spectroscopy of carbon-rich
 post-AGB stars}

\begin{abstract}

We present a detailed chemical analysis on the basis of 
high-resolution, high signal-to-noise optical spectra of two 
post-AGB objects \irt\ and \irv. Both display 
the unidentified 21 \mic\ feature in their IR-spectra. The analysis is
performed using \hra (F0Ib) and \hrz  (F5Iab) as reference stars.

The spectroscopic indicators provide accurate atmospheric parameters of  
$T_{\rm eff}$=6500 K, $\log g$=1.0 and $\xi_t$=5.5 \km\ for \irt\ and
$T_{\rm eff}$=7000 K, $\log g$=1.0 and $\xi_t$=4.0 \km\ for \irv.
Our high-resolution data are inconsistent
with the significantly lower temperatures deduced from spectral-type 
determinations in the literature based on low resolution spectra 
and highlight the need of high-resolution 
spectroscopy for the determination of accurate fundamental 
parameters of chemically peculiar supergiants.

Both photospheres are found to be metal-deficient with [Fe/H]=$-$0.4 and
$-$0.7 respectively.
C and N are found to be overabundant. Useful O-lines were only detected in
the brighter \irt\ and the O abundance is found to follow the Fe 
deficiency : the C/O photospheric abundance is about 1.3, but due to the
lack of oxygen lines it is difficult to determine accurately. This corroborates
the fact that the carriers of the 21 \mic\ feature are formed in a 
carbon rich circumstellar chemistry. 
Moreover, these IRAS-stars have large overabundances
of s-process-elements. The mean abundance of all the measured 
s-process-elements
is [s/Fe]=+1.0 for \irt\ and +1.4 for \irv. The distribution of the s-process
elements can best be described as due to a distribution of neutron exposures
with a low mean neutron exposure of $\tau_{0} = 0.2\ \rm{mbarn^{-1}}$.
The 21 \mic\ stars form an interesting sub-group in the total post-AGB sample
of stars, not only for their IR characteristics, but also in a broader
context of stellar (chemical) evolution theory. They show, in contrast to
other post-AGB stars, that the 3rd dredge-up has been efficient during
their AGB evolution. The mean neutron exposure is lower than expected
for their metallicity.

The spectroscopic parameters found for the massive spectral analogues
($T_{\rm eff}$=7500 K, $\log g$=2.0 and 
$\xi_t$=3.0 \km\ for \hra\ and $T_{\rm eff}$=6500 K, $\log g$=2.0 and
$\xi_t$=3.5 \km\ for \hrz) agree well with the values found in the
literature. Both stars display an enrichment in Na, which is commonly 
observed in massive supergiants but theoretically not well understood.

\end{abstract}

\section{Introduction}

The late AGB and post-AGB phases are among the least understood 
in current low- and intermediate-mass stellar evolution theory, mainly because
of the coarse understanding of the mass-loss mechanism(s) and 
internal convection. The theoretical understanding 
of the total chemical evolution and certainly the chemical enrichment due to
the 3rd dredge-up phenomena are therefore still poorly 
understood (e.g. Lattanzio et al. 1996).

One of the tracers of the chemical evolution of the star is the
chemical content of the dusty circumstellar envelope which is formed by the
high mass-loss episode(s) during the late AGB phase.
The infrared spectral region is often used to infer the envelope composition 
since not only the bulk of radiation is emitted 
in the IR but also because the IR-spectra are characterized 
by the chemistry. The main signature of
O-rich dusty environments are silicate features at
9.7 \mic\ and 18 \mic\ while 
a C-rich chemistry is characterized by the presence of the SiC emission at
10-12 \mic\ and/or the presence of 3.3, 6.2, 7.7 and 11.3 \mic\ features 
which are usually attributed to the polycyclic aromatic hydrocarbons (PAHs).
The spectrographs on board of ESA's Infrared Space Observatory (ISO) are
revealing now a wealth of detail concerning these 2 chemical types of
circumstellar environments around evolved stars (see e.g. the ISO dedicated 
November 1996 issue of A\&A).

Recently the detection of a strong broad emission feature at 21 \mic\ 
accompanied by unusually strong 3.4-3.5 and 6-9 \mic\ emission features 
in several post-AGB stars has been reported 
(Kwok et al. 1989; Hrivnak \& Kwok 1991a,b; Kwok et al. 1995; 
Geballe et al. 1992; Henning et al. 1996). While the carrier
of the broad feature is not yet identified 
(e.g. Begemann et al. 1996 and references therein), there are several
observational indications that it is only present in a C-rich 
circumstellar chemistry: the 21 \mic\ stars usually also display strong
PAH features at lower wavelengths; Hrivnak
(1995) studied low resolution spectra and revealed the presence of 
circumstellar C$_{2}$ and C$_{3}$ and also Bakker et al. (1995,1996,1997)
detected the
optical circumstellar carbon molecules around 21 \mic\ stars; 
Omont et al. (1993) deduced a C-rich environment from the large
HCN/CO millimetre-line ratio. Till now the 21 \mic\ feature is only observed in
post-AGB stars and young PN's and it is not clear whether the carrier is only
produced during the transition from the AGB to the PN phase or already produced
on the AGB but not excited.

Since the discovery of Justannont et al. (1996) that also 
HD~187885 may display a small
21 \mic\ feature, {\sl all} known the post-AGB stars with photospheres showing
the yields of an efficient 3rd dredge-up (Van Winckel 1997 and references 
therein), also display the 
21 \mic\ feature in their IR spectrum. The study of the stellar photospheres 
of the 21 \mic\ stars is 
therefore not only interesting for the study of the carrier and excitation
mechanism of the 21 \mic\ feature itself, but also in a broader context of
stellar (chemical) evolution of low and intermediate mass stars.

In this paper, we report on a detailed chemical study of the 
stellar photospheres of two 21 \mic\ post-AGB stars: \irt\ and \irv. 
The two objects are optically faint (V = 9.6 and 14.2 respectively) with
a total optical flux which is smaller than the total infrared flux. 
Their spectral energy distribution is double-peaked, with visible and
near-infrared components due to the reddened photosphere and a
far-infrared component caused by the detached dust shell
(Hrivnak \& Kwok 1991a). The huge IR excess may indicate a relatively high
initial mass. In the IRAS colour-colour diagram, the two stars are 
located in the region of post-AGB stars in between the late AGB 
stars and young PN's (Omont et al. 1993).
We used two bright massive supergiants \hrz\  ($\alpha$ Per) 
and \hra\  ($\alpha$ Lep) of similar spectral type 
(F5Ib and F0Ib respectively) as reference objects (see Table~\ref{model}).

\section{Observations and data reductions}

The chemical analysis of the programme stars is based on an accurate
study of high-resolution optical spectra
($\lambda$/$\delta\lambda$=50000, corresponding to a projected slit of
1.1"). The spectra have been obtained by one of us (EJB) with the Utrecht
Echelle Spectrograph, fed by the 4.2m William Herschel Telescope at
La Palma, Spain. The echelle spectrograph is equipped with a
cross-dispersor so that the different orders are projected next to one
another on the CCD. In the red
part of the spectrum the consecutive orders are not overlapping. The 
wavelength coverage of the spectra is indicated in Table~\ref{model}.
The data were reduced using the echelle package of IRAF V2.9 in a standard 
way with optimal extraction of the orders.

\section{Analysis}

\subsection{Atomic Data}

For the line identification we mainly used the line lists by
Th\'{e}venin (1989, 1990), which are based on the lines identified in
the solar spectrum by Moore et al. (1966). These lists are completed
with the line lists of Van Winckel (1995) and Bakker (1995).
We first performed a complete identification of the high S/N
spectra (S/N $>$ 250) of \hra\ and \hrz\ which we used to identify
and de-blend the lower quality spectra of the IRAS sources.

In recent years considerable efforts have been made in
calculating improved model stellar atmospheres and oscillator strengths
($\log gf$) for atomic transitions. Unfortunately, high precision
gf-values are still unavailable for many elements. We have taken the
excitation potential and the oscillator strengths mainly
from the inverted solar abundance analysis of Th\'{e}venin (1989, 1990). 
The atomic data for O~I, Mg~I, Mg~II, Si~I,
Si~II, Ti~I, Ti~II, Cr~I, Cr~II, Fe~II, Sr~I, Sr~II, Y~I, Y~II, Zr~I and Zr~II are
obtained from the automatic data bank of Kurucz
(http://cfa-www.harvard.edu/amp/data/kur23/sekur.html) and the 
mailserver of the 'Vienna Atomic Line Data Base'. 
For the Fe~I oscillator strengths we used the critical compilation  
of Lambert et al. (1996).

\subsection{Determination of the atmospheric parameters}

We used the CDROM grid of LTE model atmospheres of Kurucz (1993) in combination
with his abundance calculation programme WIDTH9. 
A model atmosphere is uniquely determined by the
metallicity, the effective temperature ($T_{eff}$), the gravity
($\log g$) and the microturbulent velocity ($\xi_{t}$). We used input 
models with a solar metallicity Z for \hra\ and \hrz\ and Z=$-$0.5 for the
IRAS-sources with $\xi_t$=2 \km.

The determination of these atmospheric parameters were solely based 
on our high-resolution spectra. Quantitative photometric analysis is hampered
by the uncertainty on the amount of circumstellar reddening, the possible
anomalous circumstellar reddening law and by the inaccurate 
calibration of the photometric systems for supergiants in general. 
The photometry can therefore only be used as a first guess for the finer 
spectroscopic analysis.

\subsubsection{The effective temperature}

A spectroscopic estimate for
$T_{eff}$ is found by forcing the abundances of an ion 
to be independent on the lower excitation potential of the transitions. 
For F-type supergiants, only Fe~I lines are suitable for such an 
analysis since 
this method is only usable for ions with a large number of weak
lines having a large spread in excitation potential. Moreover, for 
supergiants in
the temperature range of our programme stars, non-LTE effects on the Fe~I
lines are small (Venn 1995b).

During our research it became clear that even a few unprecise oscillator
strength values can limit the precision of this method considerably. 
When using the Fe~I oscillator strengths of Th\'evenin we encountered
the same problem for every star : the effective temperature determined
on the base of the Fe~I lines was some 500K higher than the temperature 
based on the
Fe~II lines. Moreover, the temperature estimates of the reference stars
were higher than the temperatures found in the literature.

First we situated this discrepancy in the departure
of local thermodynamic equilibrium (non-LTE  effects). The LTE underabundance 
of
iron due to overionisation of Fe~I in F supergiants amounts from $\sim$0.03 dex
(F8 supergiants) to 0.2 dex (F0 supergiants) (Venn 1995b; Boyarchuk \& 
Lyubimkov 1983; Boyarchuk et al. 1985, 1988) A shift of 0.2 dex in abundance
corresponds to a temperature shift of $\sim$200 K. It turned out, however,  
that for every star $\Delta\/T_{\rm eff}$
(=$T_{\rm eff}$ (Fe~I)-$T_{\rm eff}$ (Fe~II)) was 
too large (e.g. 700 K for \hra\ and 400 K for \hrz) to be caused solely by
non-LTE effects.  
Note that by limiting our abundance analysis to lines with an 
equivalent width smaller than 150 m\ang, we
focus on lines formed in deeper layers where the non-LTE effects are known to
be smaller. 

\begin{figure}
\begin{center}
\mbox{\epsfxsize=0.48\textwidth\epsfysize=5.5cm\epsfbox[30 45 545 460]
{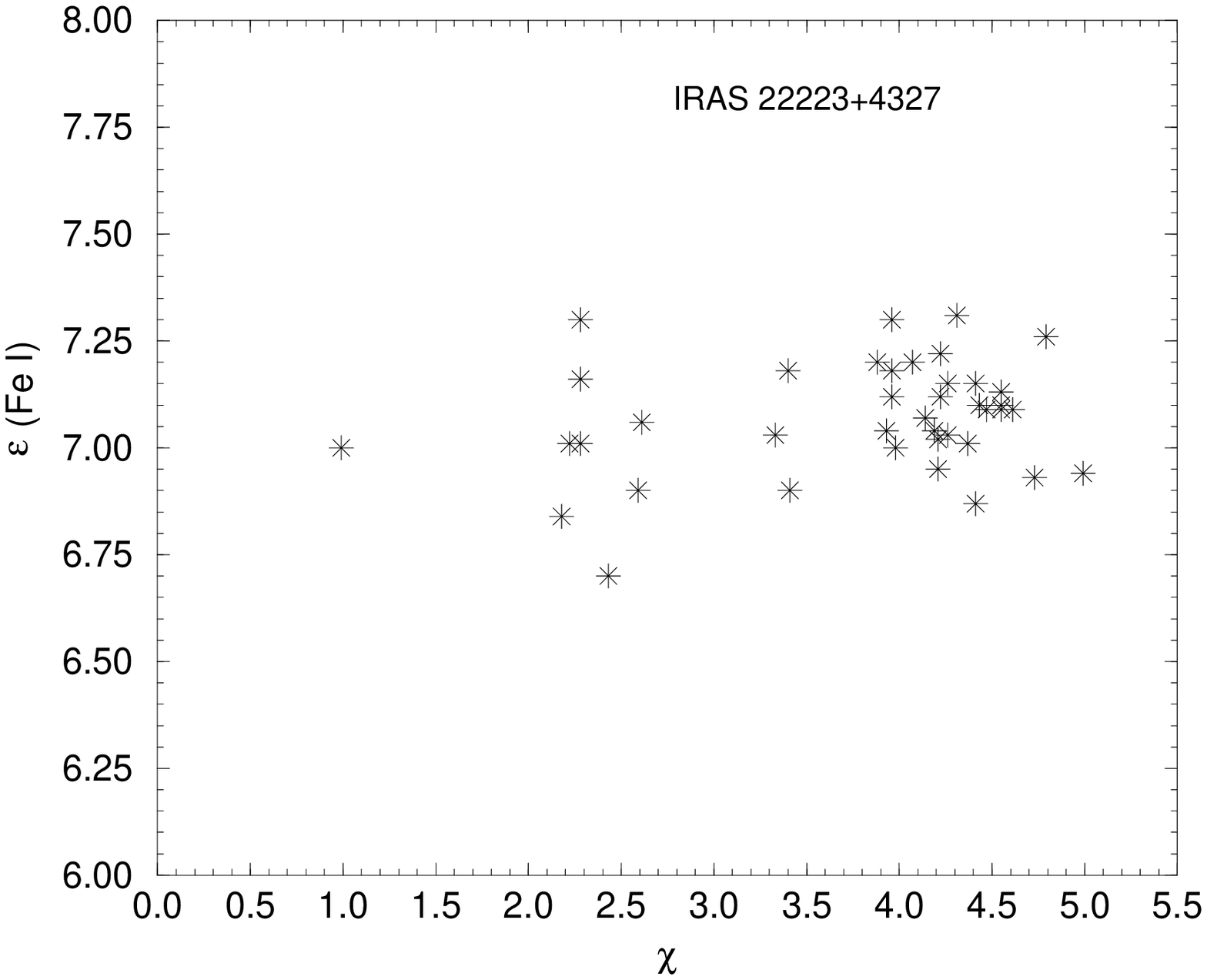}}
\mbox{\epsfxsize=0.48\textwidth\epsfysize=5.5cm\epsfbox[30 45 545 460]
{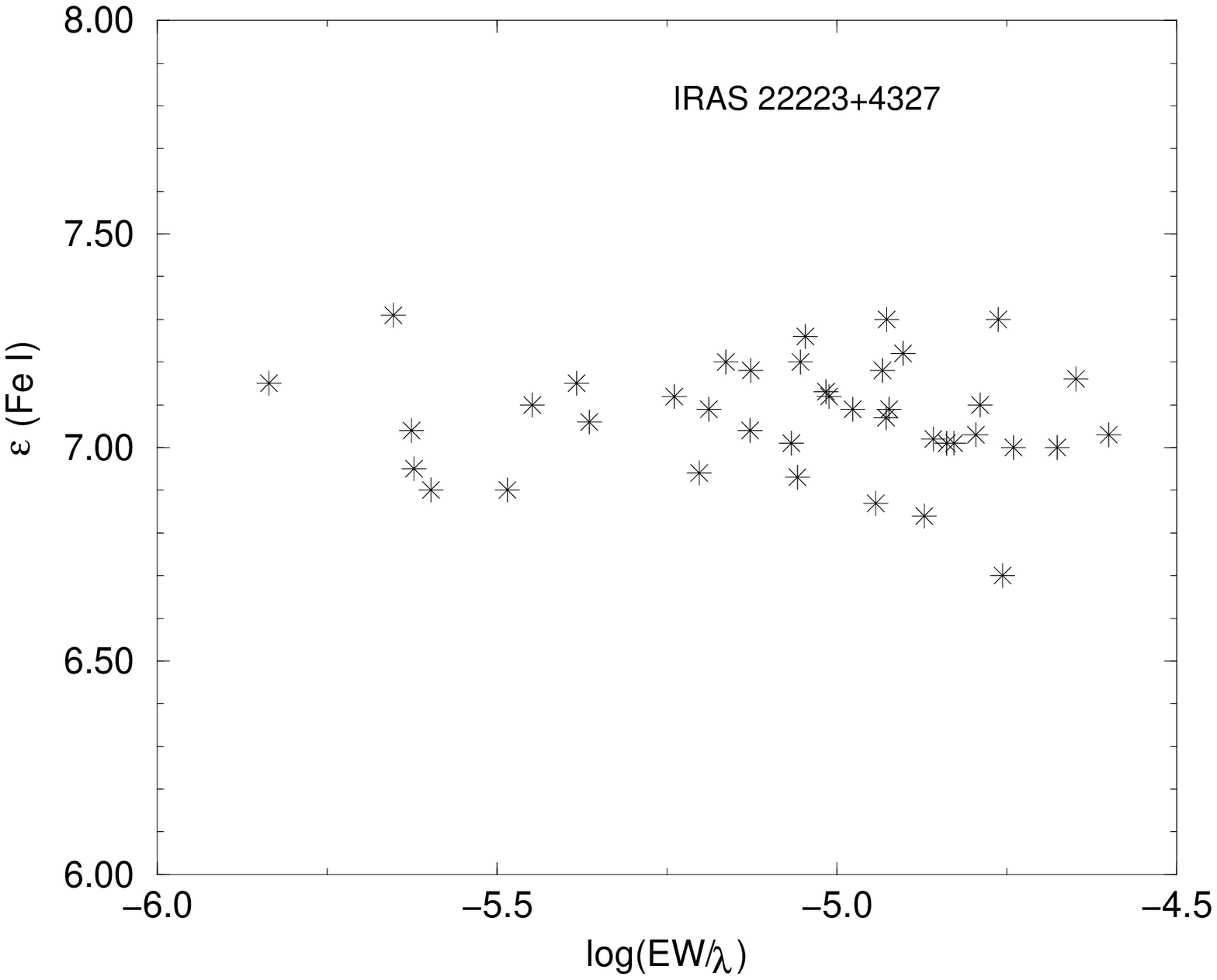}}
\end{center}
\caption{\label{tempirt}The excitation potential-abundance
diagram and the equivalent width-abundance diagram 
for IRAS~22223+4327.}
\end{figure}

We have then limited the Fe~I lines to only those listed
in the critical compilation of $\log gf$-values by Lambert et al. (1996). 
They re-discussed the Fe~I 
gf-values by comparing the gf-values of several authors (see Lambert for
references). By using these oscillator strengths, all problems concerning
discrepancy of $T_{\rm eff}$, $\log g$ or $\xi_t$ (non-LTE effects)
disappear! It is
extremely important to  limit the used Fe~I lines to transitions with a precise
oscillator strength. For 
Fe~II we don't have any lines in common with Lambert. Therefore we have 
compared
the $\log gf$-values of Th\'evenin with these of Lambert: the $\log gf$-values
of Th\'evenin were systematically 0.19$\pm$0.08 smaller than these of Lambert,
due to the fact that Lambert used 7.51 for the solar Fe abundance and 
Th\'evenin
7.67. We have thus increased the $\log gf$-values of Fe~II of
Th\'evenin with 0.19. The solar abundances of the other chemical elements are
from Grevesse (1989).
For \hra, \hrz\  and \irt\  we derived a temperature of 7500 K, 6500 K and
6500 K respectively (e.g. see Fig. \ref{tempirt}).

\begin{figure}
\begin{center}
\mbox{\epsfxsize=0.48\textwidth\epsfysize=8cm
\epsfbox[135 130 550 780]{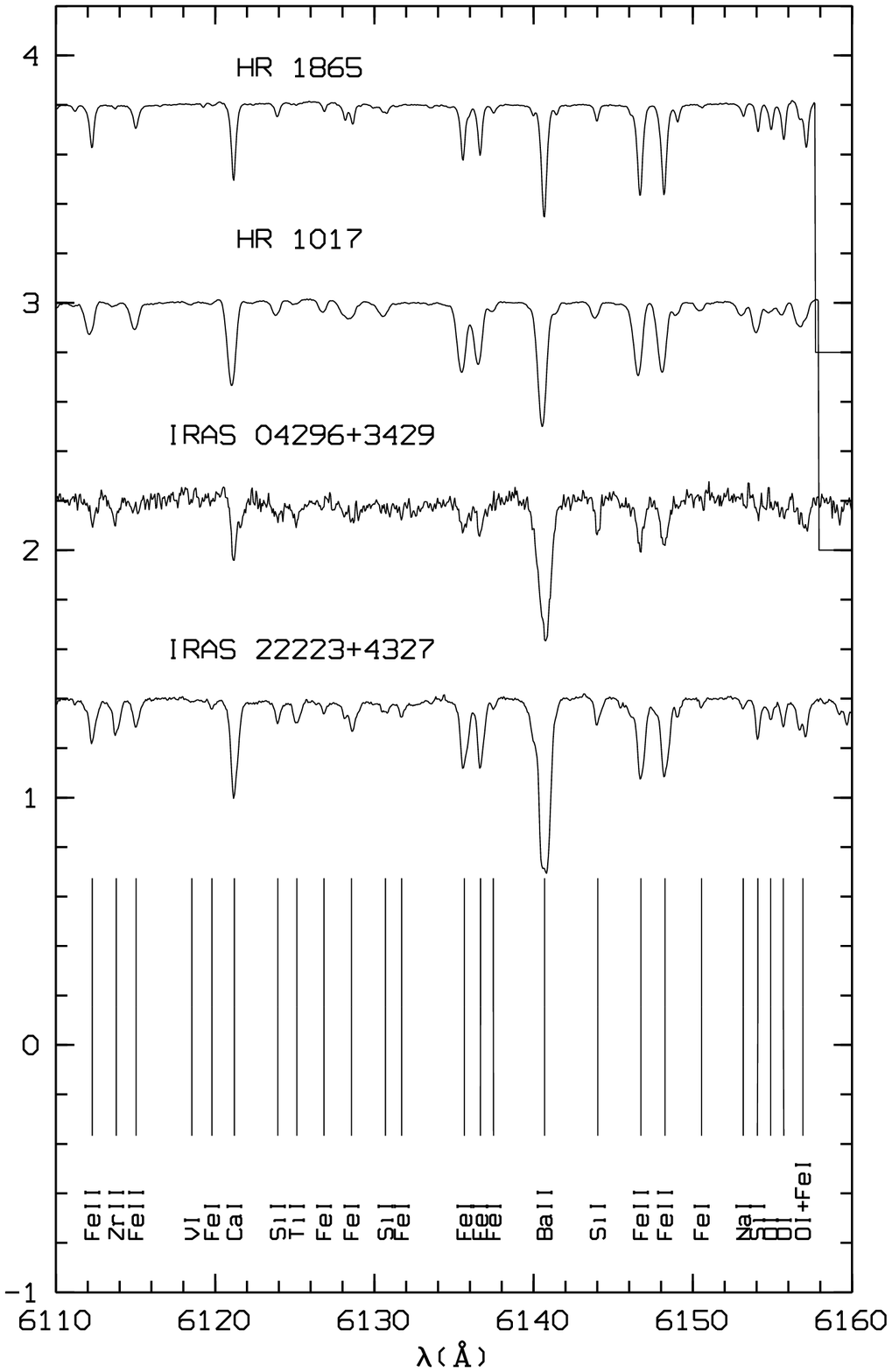}}
\caption{\label{spectra}Spectra of \hra, \hrz, \irt\ and \irv.}
\end{center}
\end{figure}

For the faint object \irv\, it turned out to be impossible to
determine the effective temperature with this method due to the
lack of good quality Fe~I lines. A rough estimate is found 
when comparing the spectra of the IRAS-stars with those of the two 
reference stars. Figure \ref{spectra} shows
the spectra of the four programme stars in the wavelength coverage between 6110
\ang\ and 6160 \ang. This is an interesting spectral interval since besides
the O~I triplet and Ba~II line, also Fe~I and Fe~II lines are present.
The ratio of the Fe~II line-strengths versus the Fe~I lines gives an 
indicator of the effective temperature for these four stars of similar 
gravity.

The resemblance between
the spectra of \hrz\ and \irt\ is striking. This is 
confirmed by our spectroscopic temperature assessment of
$T_{\rm eff}$=6500 K for \hrz\ and 
\irt. For \irv\ the ratio of the Fe~II to Fe~I lines is higher than the
ratio of \hrz\ and lower than this of \hra. We therefore estimate the
temperature of \irv\ to be approximately 7000 K. To test the consistency of
these $T_{\rm eff}$, we have checked the excitation 
potential-abundance diagram of \irv. The
slope is small and positive, but it indicates that the temperature departure
is still less than 300 K, within the bounds of the estimated accuracy.

Hrivnak (1995) has determined the spectral type of the IRAS sources 
on the basis of low resolution spectra and concluded that both
 \irv\ and \irt\ are G0Ia supergiants. This spectral classification was used 
by Kwok et  al. (1995) and Bakker et al. (1996) to estimate the effective 
temperature 
of 5000-5500 K for both stars. This low temperature is, however,
inconsistent with our high-resolution spectra! Not only because
in the lower excitation potential-abundance plot, a significant upward trend
shows up, 
but also because neutral lines of heavy elements should appear in the
spectra at those temperatures. The strong Y I line at 6435 \ang\ for 
instance, would have an equivalent width of 20 m\ang\ for
the abundance computed from the Y II lines at a temperature of 5500 K
and log(g) = 1.0.

We want to stress the fact that for reddened, chemically peculiar 
supergiants, high-resolution data are definitely needed for accurate 
fundamental parameter determinations. The line regions, used in the low 
resolution spectra to infer the
spectral type, are influenced by strong resonance lines of s-process elements
like Ba and Y. With strong s-process enhancements, like observed in 
IRAS~05341+0852 by Reddy et al. (1997), the optical spectrum is even 
completely dominated by lines of s-process isotopes. In our opinion, the
chemical peculiarity and certainly the strong enhancements of s-process 
isotopes, make standard spectral classification difficult.

\subsubsection{Gravity and the microturbulent velocity}

We derived the model gravity by implying that different ions of
the same element yield the same abundances to within 0.1 dex.
For F supergiants, the only element useful for this purpose is again Fe 
by the lack of alternatives with enough lines of both ionisation stages. 
The observed hydrogen Balmer lines are affected by emission so we did not use
them to constrain the gravity. A change in the $\log g$ value by 0.5 
makes the abundances derived for the two ionisation stages differ by in between
0.11 and 0.26 dex.
The total uncertainty on the gravity of these objects is, however,
more uncertain than the $\pm$ 0.3 this method suggests, 
since it does not take into account other 
uncertainties on the abundances which are more difficult to quantify, 
like non-LTE effects, systematic $\log gf$ errors etc. 
Note that the main conclusions of this work (see discussion)
are based on abundance {\sl ratios}, which are much less influenced by 
uncertainties in effective temperature and gravity than absolute values.

The microturbulent velocity for each star has been determined
by forcing the Fe~I abundances to be independent on the equivalent
width (W$_{\lambda}$) (see Fig. \ref{tempirt} for \irt).

\irv\  has not enough lines available with a good range in equivalent
width for a reliable estimate. We assumed $\xi_t$=4.0 \km. Nevertheless, the
value of microturbulence is not critical in this analysis since only 
weak lines (W$_{\lambda}\le$150 m\ang) are included in the final average
abundances. The use of weak lines also means that radiative damping terms will
not be important.

A synopsis of the atmospheric parameters for the four stars
is listed in Table~\ref{model}.

\tabcolsep=4pt
\begin{table}[!ht]
\caption{\label{model}The atmosphere models for \hra, \hrz, \irt\ and \irv.}
\vspace{0.5ex}
\begin{center}
\scriptsize
\begin{tabular}{|c|ccccc|}\hline
\rule[-3mm]{0mm}{8mm} &  m(v) & $T_{eff}$ & $\log g$ & $\xi_t$ & Obs. region\\
                    & magn. & \footnotesize{K} &
 \footnotesize{$\rm{cm\, s^{-2}}$} & 
\footnotesize{\km } & \ang \\
\hline
\rule[-0mm]{0mm}{5mm} IRAS  & & & & & \\
\rule[-3mm]{0mm}{3mm} 04296+3429 & {\raisebox{1.5ex}[0pt]{14.2}} &
{\raisebox{1.5ex}[0pt]{7000}} & {\raisebox{1.5ex}[0pt]{1.0}} &
{\raisebox{1.5ex}[0pt]{4.0}} & {\raisebox{1.5ex}[0pt]{5560--10045}}\\
\rule[-0mm]{0mm}{5mm} IRAS  & &  & & & \\
\rule[-3mm]{0mm}{3mm} 22223+4327 & {\raisebox{1.5ex}[0pt]{9.7}} &
{\raisebox{1.5ex}[0pt]{6500}} & {\raisebox{1.5ex}[0pt]{1.0}} &
{\raisebox{1.5ex}[0pt]{5.5}} &{\raisebox{1.5ex}[0pt]{4430-10045}} \\
\rule[-3mm]{0mm}{8mm} HR~1017  & 1.7 & 6500 & 2.0 & 3.5 &
4690--7340 \\
\rule[-3mm]{0mm}{8mm} HR~1865  & 2.6 & 7500 & 2.0 & 3.5 &
4690--7340 \\
\hline
\end{tabular}
\end{center}
\end{table}
\normalsize

\subsection{Error analysis}

\subsubsection{Internal errors}

The standard deviation $\sigma$ on the abundance of the element for which we
observed more than five lines is a
good indicator of the consistency of the chemical analysis (e.g. see
Table~\ref{abhr8}). A typical value for $\sigma$ is between 0.10 and 0.25 dex.

For a good model in ionisation and excitation equilibrium, $\sigma$ is
mainly determined by non-systematic errors on the equivalent width and
especially on the oscillator strength.
The equivalent widths were measured by fitting a Gaussian curve to
the absorption lines.

To check our $\log gf$-values for systematic errors, we compared our 
LTE abundance analysis of \hra\ with a similar study of Venn (1995b). 
Venn mainly used the oscillator strengths of
F\"uhr et al. (1988); Wiese \& Martin (1980); F\"uhr et al. (1981);
Wiese \& Fuhr (1975) and O' Brian et al. (1991). No systematic differences
occur, so we continued to use the $\log gf$-values of Th\'evenin for most 
atomic species.
Since we did not account for the hyperfine broadening
of transitions of atoms with an odd atomic number, the abundances
of these ions are slightly overestimated.

Other contributors to the scatter could be differential non-LTE effects
and non-detected blends. 
The blends of the IRAS stars, which have broader
lines, are detected by comparing the spectra with the narrow lined reference
stars.

\subsubsection{Inaccuracies on the model parameters}

The model parameters
($T_{eff}$, $\log g$ and $\xi_t$) are not independent: a change in one
parameter generally induces a shift in another for the
spectroscopic requirements (ionisation balance, independence of the
abundance of an ion versus the excitation potential and 
equivalent width) to be fulfilled. A typical shift of 0.5 dex in the
gravity induces a temperature shift of 300-400 K, so that the
ionisation balance still would be maintained. We refer to Table~3 of
Van Winckel (1997) for a quantitative estimation of the influence of
the different uncertainties on the abundance determination. The
temperature uncertainty has by far the biggest influence on the
abundance accuracy, especially for ions with the smallest occupation level
since they are most influenced by uncertainties on the ionisation balance
induced by errors on the temperature and gravity.

\section{Results of the chemical analyses}

In the Tables~\ref{abhr8}-\ref{abir4} we give
the results of the abundance analysis for each programme star. For every ion we
list the number of 
lines used, the mean equivalent width, the absolute abundance, the abundance
ratio relative to the solar value  and the internal scatter, if more
than one line is used. A complete line list with the detailed atomic
data can be obtained upon request. For the solar iron abundance we used the
meteoric iron abundance of 7.51.

\begin{table}[!ht]
\caption{\label{abhr8}Chemical analysis of HR~1865. In the last column we
listed the difference with the analysis of Venn (1995b).}
\vspace{0.5ex}
\begin{center}
\begin{footnotesize}
\begin{tabular}{|lrr|llll|}\hline
\multicolumn{7}{|c|}{\rule[-0mm]{0mm}{5mm}{\bf HR~1865}}\\
\multicolumn{7}{|c|}{{\em $T_{eff}$=7500 K}}\\
\multicolumn{7}{|c|}{{\em $\log g$=2.0}}\\
\multicolumn{7}{|c|}{{\em $\xi_t$=3.0 \km}}\\
\multicolumn{7}{|c|}{\rule[-3mm]{0mm}{3mm} {\em [Fe/H]=0.00}}\\
\hline
\rule[-3mm]{0mm}{8mm}ion & N & $\overline{W_{\lambda}}$ & $\epsilon$ & [el/Fe] 
& $\sigma$ & $\Delta$Venn\\
\hline
\rule[-0mm]{0mm}{5mm}C~I  & 11 & 29 & 8.16 & $-$0.40  & 0.09 & $+$0.02 \\
O~I  & 5  & 34 & 8.44 & $-$0.49 & 0.02   & $-$0.24 \\
Na~I & 3  & 67 & 6.72 & +0.39  & 0.11   & $-$0.20 \\
Mg~I & 2  & 23 & 7.52 & $-$0.06  & 0.05   & $-$0.26 \\
Si~I & 9  & 19 & 7.34 & $-$0.21 & 0.26  & $-$0.08 \\
S~I  & 3  & 29 & 7.22 & +0.01  & 0.06    & $-$0.22 \\
Ca~I & 9  & 9 & 6.44 & +0.08  & 0.13    & $-$0.06 \\
Sc~II& 1  & 14 & 3.27 & +0.17  & \     & $+$0.23 \\
Ti~I & 4  & 16 & 4.97 & $-$0.02  & 0.07   &  \\
Ti~II& 2  & 101 & 5.07 & +0.08  & 0.09 & $+$0.19 \\
Cr~I & 2  & 15 & 5.57 & $-$0.10  & 0.05    &  \\
Cr~II& 4  & 90 & 5.88 & +0.21  & 0.12   & $+$0.21 \\
Mn~I & 3  & 17 & 5.46 & +0.07  & 0.13    & \\
Fe~I & 59 & 36 & 7.50 & $-$0.01 & 0.12  & $-$0.15 \\
Fe~II& 18 & 50 & 7.53 & $+$0.02  & 0.15  & $+$0.10 \\
Ni~I & 28 & 17 & 6.18 & $-$0.07  & 0.14   & $-$0.08 \\
Cu~I & 1  & 5 & 3.98 & $-$0.23  & \      & \\
Zn~I & 1 & 17 & 4.36 & $-$0.24  & \     & \\
Y~II & 8 & 47 & 2.41 & +0.17  & 0.16    &  \\
Zr~II & 5 & 18 & 2.98 & +0.38  & 0.21   & $+$0.19 \\
Ce~II & 5 & 7 & 1.71 & +0.16 & 0.09     &   \\
\rule[-3mm]{0mm}{3mm}Nd~II & 7 & 9.2 & 1.73 & +0.23 & 0.10   &  \\
\hline
\end{tabular}
\end{footnotesize}
\end{center}
\end{table}

\begin{table}[!ht]
\caption{\label{abhr7}Chemical analysis of HR~1017.} 
\vspace{0.5ex}
\begin{center}
\begin{footnotesize}
\begin{tabular}{|lrr|lll|}\hline
\multicolumn{6}{|c|}{\rule[-0mm]{0mm}{5mm} {\bf HR~1017}}\\
\multicolumn{6}{|c|}{ {\em $T_{eff}$=6500 K}}\\
\multicolumn{6}{|c|}{ {\em $\log g$=2.0}}\\
\multicolumn{6}{|c|}{{\em $\xi_t=3.50$ \km}}\\ 
\multicolumn{6}{|c|}{\rule[-3mm]{0mm}{3mm} {\em [Fe/H]=$-$0.16}}\\ 
\hline
\rule[-3mm]{0mm}{8mm}ion & N & $\overline{W_{\lambda}}$ & $\epsilon$ & [el/Fe] 
& $\sigma$ \\
\hline
\rule[-0mm]{0mm}{5mm}C~I & 10 & 47 & 8.43 & $+$0.03 & 0.11 \\
O~I & 2 & 32 & 8.65 & $-$0.12 & 0.14 \\
Na~I & 1 & 91 & 7.01 & $+$0.84 & \  \\
Al~I & 1 & 25 & 6.31 & $+$0.00 & \  \\
Si~I & 8 & 33 & 7.19 & $-$0.20  & 0.22 \\
Ca~I & 7 & 78 & 6.24 & $+$0.04 & 0.11 \\
Sc~II & 3 & 43 & 3.19 & $+$0.25 & 0.06 \\
Ti~I & 7 & 16 & 4.84 & $+$0.01 & 0.14 \\
Ti~II & 4 & 94 & 4.83 & $+$0.00 & 0.12 \\
Cr~I & 7 & 44 & 5.50 & $-$0.01 & 0.18 \\
Cr~II & 7 & 87 & 5.75 & $+$0.24 & 0.13 \\
Mn~I & 6 & 26 & 5.32 & $+$0.09 & 0.16 \\
Fe~I & 69 & 44 & 7.32 & $-$0.03 & 0.12 \\
Fe~II & 12 & 47 & 7.39 & $+$0.04 & 0.07 \\
Ni~I & 30 & 36 & 6.22 & $+$0.13 & 0.13 \\
Cu~I & 1 & 36 & 3.98 & $-$0.07 & \  \\
Zn~I & 1 & 38 & 4.03 & $-$0.41 & \  \\
Y~II & 3 & 37 & 2.01 & $-$0.07 & 0.11 \\
Zr~II & 2 & 33 & 2.64 & $+$0.20 & 0.25 \\
La~II & 2 & 20 & 1.44 & $+$0.38 & 0.04 \\
Ce~II & 5 & 17 & 1.37 & $-$0.02 & 0.06 \\
\rule[-3mm]{0mm}{3mm}Nd~II & 15 & 23 & 1.39 & $+$0.05 & 0.14 \\
\hline
\end{tabular}
\end{footnotesize}
\end{center}
\end{table}

\begin{table}[!ht]
\caption{\label{abir2}Chemical analysis of IRAS~22223+4327.}
\vspace{0.5ex}
\begin{center}
\begin{footnotesize}
\begin{tabular}{|lrr|lll|}\hline
\multicolumn{6}{|c|}{\rule[-0mm]{0mm}{5mm} {\bf IRAS~22223+4327}}\\
\multicolumn{6}{|c|}{ {\em $T_{eff}$=6500 K}}\\
\multicolumn{6}{|c|}{ {\em $\log g$=1.0}}\\
\multicolumn{6}{|c|}{{\em $\xi_t=5.50$ \km}}\\ 
\multicolumn{6}{|c|}{\rule[-3mm]{0mm}{3mm} {\em [Fe/H]=$-$0.44}}\\ 
\hline
\rule[-3mm]{0mm}{8mm}ion & N & $\overline{W_{\lambda}}$ & $\epsilon$ & [el/Fe] 
& $\sigma$ \\
\hline
\rule[-0mm]{0mm}{5mm}C~I & 8 & 70 & 8.63 & $+$0.51 & 0.14 \\
N~I & 4 & 54 & 7.88 & $+$0.27 & 0.05 \\ 
O~I & 3 & 24 & 8.50 & $+$0.01 & 0.04  \\ 
Al~I & 4 & 16 & 6.15 & $+$0.12 & 0.10 \\ 
Si~I & 14 & 46 & 7.43 & $+$0.32 & 0.17  \\ 
S~I & 4 & 53 & 6.85 & $+$0.08 & 0.09 \\ 
Ca~I & 6 & 58 & 5.97 & $+$0.05 & 0.08 \\ 
Ca~II & 1 & 6 & 5.80 & $-$0.12 & \  \\
Ti~I & 2 & 26 & 4.95 & $+$0.40 & 0.16 \\
Cr~I & 7 & 45 & 5.19 & $-$0.04 & 0.11 \\
Cr~II & 6 & 118 & 5.39 & $+$0.16 & 0.22 \\
Mn~I & 1 & 38 & 4.92 & $-$0.03 & \\
Mn~II & 1 & 50 & 5.20 & $+$0.25 & \\
Fe~I & 41 & 57 & 7.08 & $+$0.01 & 0.13 \\
Fe~II & 12 & 49 & 7.05 & $-$0.02 & 0.15 \\
Ni~I & 19 & 35 & 5.95 & $+$0.14 & 0.19 \\
Y~II & 4 & 128 & 3.44 & $+$1.64 & 0.09 \\
Zr~II & 6 & 56 & 3.47 & $+$1.31 & 0.15 \\
Ba~II & 1 & 13 & 2.67 & $+$0.98 & \\
La~II & 16 & 68 & 1.96 & $+$1.18 & 0.17 \\
Ce~II & 8 & 52 & 2.06 & $+$0.95 & 0.23 \\
Pr~II & 5 & 47 & 1.32 & $+$1.05 & 0.24 \\
Nd~II & 11 & 51 & 1.92 & $+$0.86 & 0.22 \\
Sm~II & 10 & 54 & 1.24 & $+$0.68 & 0.25 \\
Eu~II & 2 & 51 & 0.77 & $+$0.70 & 0.06\\
&&&&& \\ \hline
\end{tabular}
\end{footnotesize}
\end{center}
\end{table}

\begin{table}[!ht]
\caption{\label{abir4}Chemical analysis of IRAS~04296+3429.}
\vspace{0.5ex}
\begin{center}
\begin{footnotesize}
\begin{tabular}{|lrr|lll|}\hline
\multicolumn{6}{|c|}{\rule[-0mm]{0mm}{5mm} {\bf IRAS~04296+3429}}\\
\multicolumn{6}{|c|}{ {\em $T_{eff}$=7000 K}}\\
\multicolumn{6}{|c|}{ {\em $\log g$=1.0}}\\
\multicolumn{6}{|c|}{{\em $\xi_t=4.0$ \km}}\\ 
\multicolumn{6}{|c|}{\rule[-3mm]{0mm}{3mm} {\em [Fe/H]=$-$0.69}}\\ 
\hline
\rule[-3mm]{0mm}{8mm}ion & N & $\overline{W_{\lambda}}$ & $\epsilon$ & [el/Fe] 
& $\sigma$ \\
\hline
\rule[-0mm]{0mm}{5mm}C~I & 7 & 87 & 8.81 & $+$0.94 & 0.18 \\
N~I & 2 & 65 & 7.84 & $+$0.48 & 0.02 \\
Si~I & 3 & 55 & 7.46 & $+$0.60 & 0.07 \\
S~I & 1 & 57 & 7.13  & $+$0.61 & \\
Ca~I & 3 & 35 & 5.93 & $+$0.26 & 0.16 \\
Fe~I & 8 & 31 & 6.85 & $+$0.03 & 0.11 \\
Fe~II & 10 & 74 & 6.80 & $-$0.02 & 0.12 \\
Ni~I & 3 & 47 & 6.12 & $+$0.56 & 0.09 \\
Y~II & 3 & 84 & 3.60 & +2.05 & 0.06 \\
Zr~II & 4 & 33 & 3.41 & +1.50 & 0.11 \\
Ba~II & 1 & 25 & 3.24 & +1.80 & \\
La~II & 5 & 78 & 2.38 & +1.85 & 0.14 \\
Ce~II & 2 & 30 & 2.14 & +1.28 & 0.05 \\
Nd~II & 2 & 57 & 1.96 & +1.13 & 0.16 \\
\rule[-3mm]{0mm}{3mm}Eu~II & 1 & 45 & 1.04 & +0.35 & \\
\hline
\end{tabular}
\end{footnotesize}
\end{center}
\end{table}

\subsection{HR~1865 and HR~1017}

Although our primary goal was to use the spectra of the 
narrow-lined supergiants \hra\ and \hrz\ for line-identification purposes
and to recognize possible blending in the spectra of the IRAS objects, 
we also performed a complete LTE abundance analysis of these two bright objects.

The non-variable HR~1865 is a very bright Galactic F-type supergiant 
which is often used as a standard star for Galactic and extragalactic
research. The most recent extensive chemical study is by Venn (1995a,b).
She deduced model-parameters of $T_{\rm eff}$ = 7400 and log(g) = 1.1 and
$\xi_t$ = 4.0 which are similar to our findings. The difference between the 
two studies is given in Table~\ref{abhr8}. Generally the agreement is
rather good with the noticeable exception of Mg~I, S~I and Sc~II where the
difference amounts to 0.26, 0.22 and 0.23 dex respectively. There are no Sc~II
and Mg~I lines 
common in our analyses. The common S~I lines (3 in total) have very
similar equivalent widths, so the difference in results is due to slightly
different atomic parameters and microturbulent velocity. 

The most recent analysis of \hrz\ we found in the literature is by
Luck \& Lambert (1985). The atmospheric parameters they found are somewhat
cooler than ours : $T_{\rm eff}$ = 6250, log(g)=0.90 and $\xi_t$ = 3.0.
Their iron abundance is somewhat higher, but generally the agreement between the
abundances is good.

\begin{figure}
\begin{center}
\mbox{\epsfxsize=0.48\textwidth\epsfbox{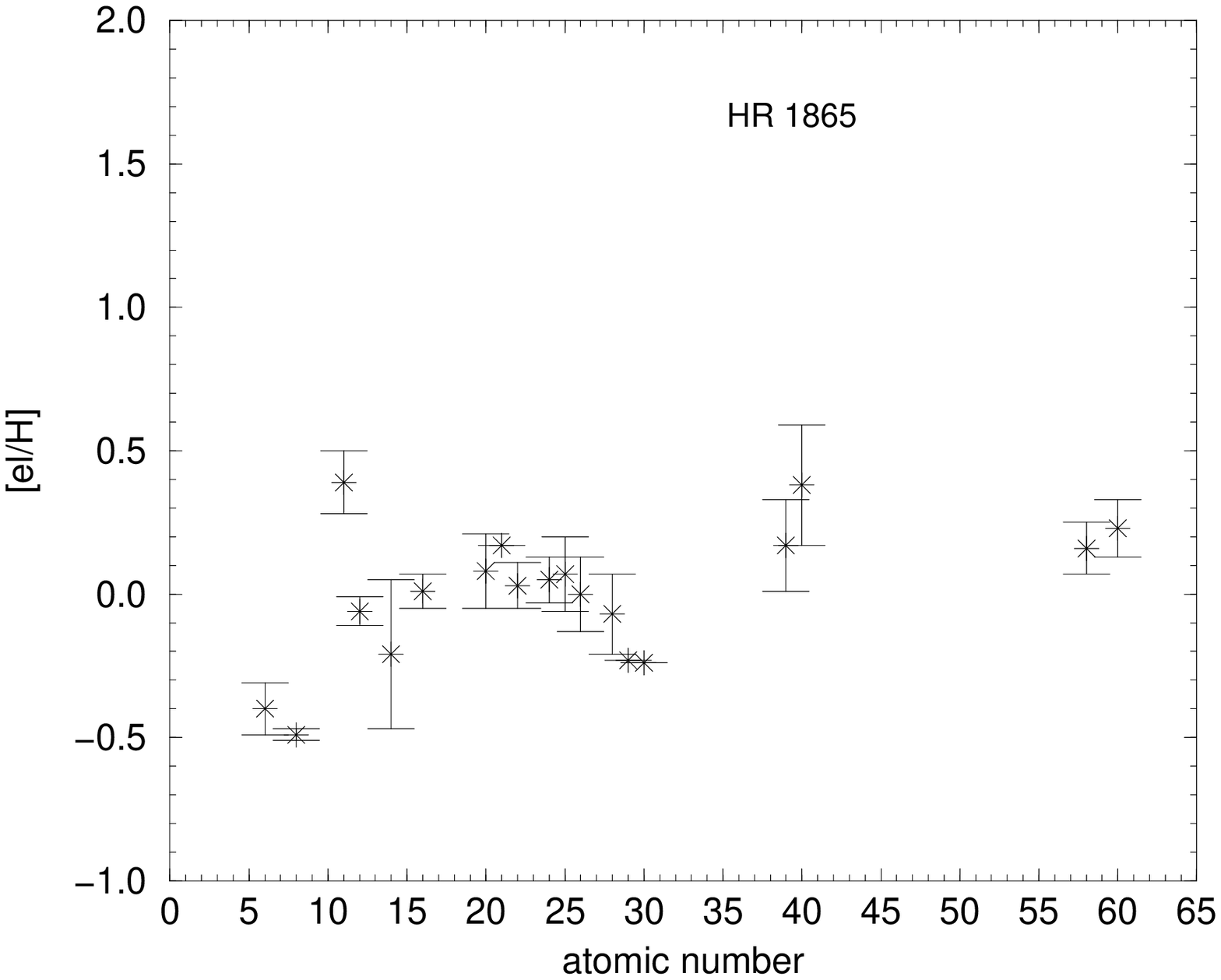}}
\mbox{\epsfxsize=0.48\textwidth\epsfbox{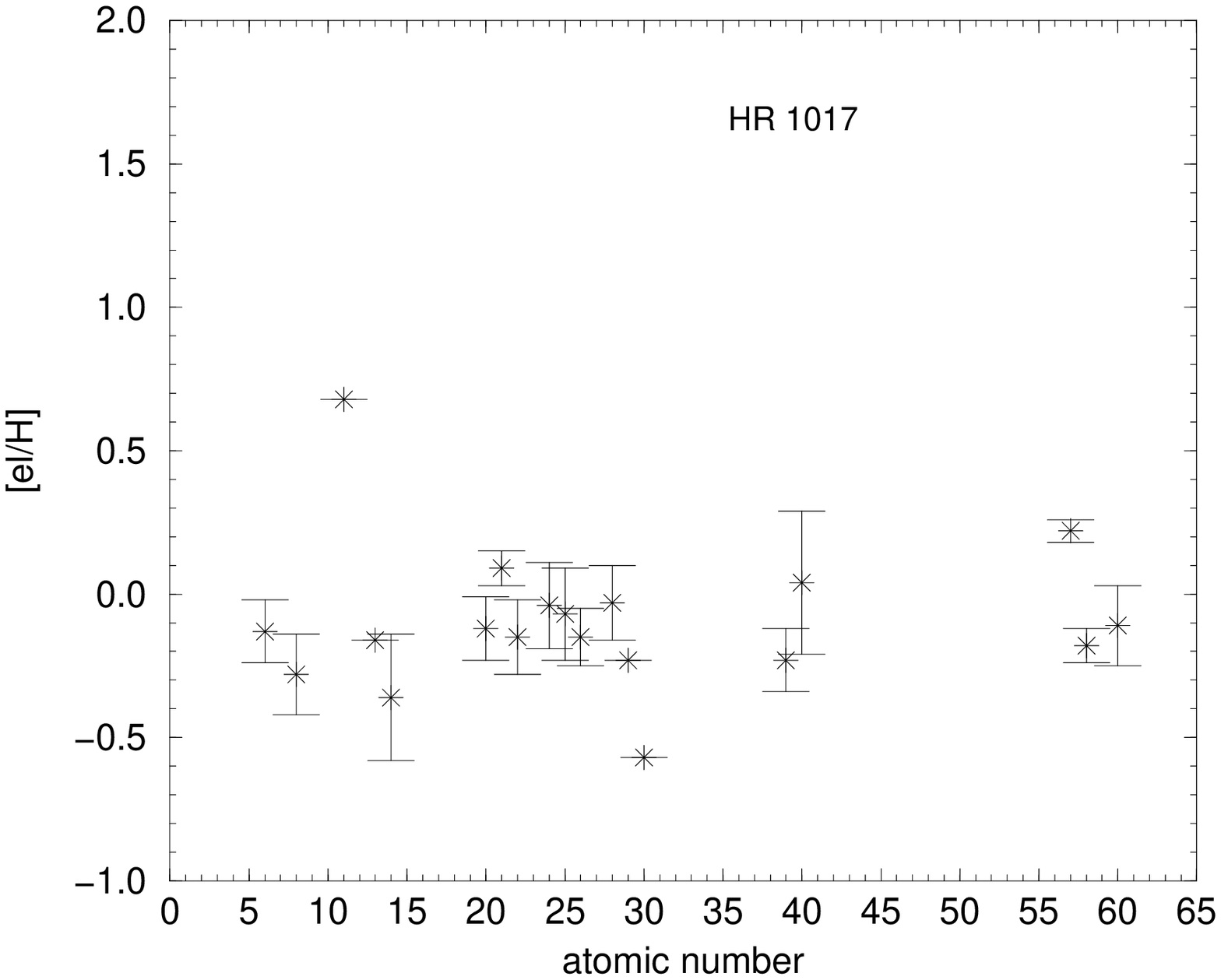}}
\caption{\label{hr1865}The abundances of HR~1865 and HR~1017
relative to the solar value.}
\end{center}
\end{figure}

\subsubsection{Intermediate and heavy elements (Na-Zr)}

Except for Na, the other elements do not seem to be enhanced (see
Fig. \ref{hr1865}). This corresponds to the abundance
analysis of Venn and Luck \& Lambert.

Both \hra\  and \hrz\  display a strong sodium enrichment ([Na/H]=+0.39 and
+0.68). Sodium enrichment in F-K supergiants was first reported by Luck (1977,
1978). Since then several authors have detected an enhancement of Na (see Venn
1995b and references therein). Boyarchuk \& Lyubimkov (1983) proposed two
possibilities for the Na-enrichment: 1) non-LTE effects in the tenuous and cool
atmosphere, of which the corrections in the abundances are $\le$ $-$0.1 dex for
weak lines 2) during a NeNa proton capture reaction
($^{22}$Ne(p,$\gamma$)$^{23}$Na) Na could have been 
synthesized and after deep mixing with the interior (first dredge-up),
Na-enrichment may occur in the stellar atmosphere (Venn 1995b). The
Na-synthesis can only take place when the temperature is high enough. Therefore
Na-enrichment is mass dependent: the more massive the star, the stronger Na
will be enhanced. Takeda \& Takada-Hidai (1994) calculated that \hra\  is a
less  massive star than \hrz\  (M$_{HR~1017}$=17 \zm\ 
and M$_{HR~1865}$=14.5 \zm); the
Na enrichment displays the same trend.

Interestingly in an LTE analysis of several high-luminosity LMC/SMC Cepheids,
Hill et al. (1995) did not found a general Na enhancement. Since the same 
temperatures 
and densities apply for the LMC/SMC Cepheid models,
there is a hint that the Na enhancement of the Galactic 
Cepheids is real and not due to non-LTE effects. Detailed evolutionary
models coupled with accurate nucleosynthetic networks 
are, however, needed in order to quantify the 
dependance of the nucleosynthetic yields and mixing ratios on the overall
chemical composition, or other as yet unexplained difference in the LMC/SMC and
Galactic supergiants.

\subsection{IRAS~22223+4327 and IRAS~04296+3429}

In Tables~\ref{abir2} and~\ref{abir4} we give an overview of the 
abundances relative to the solar value for \irt\ and \irv. 

\begin{figure}
\begin{center}
\mbox{\epsfxsize=0.48\textwidth
\epsfbox{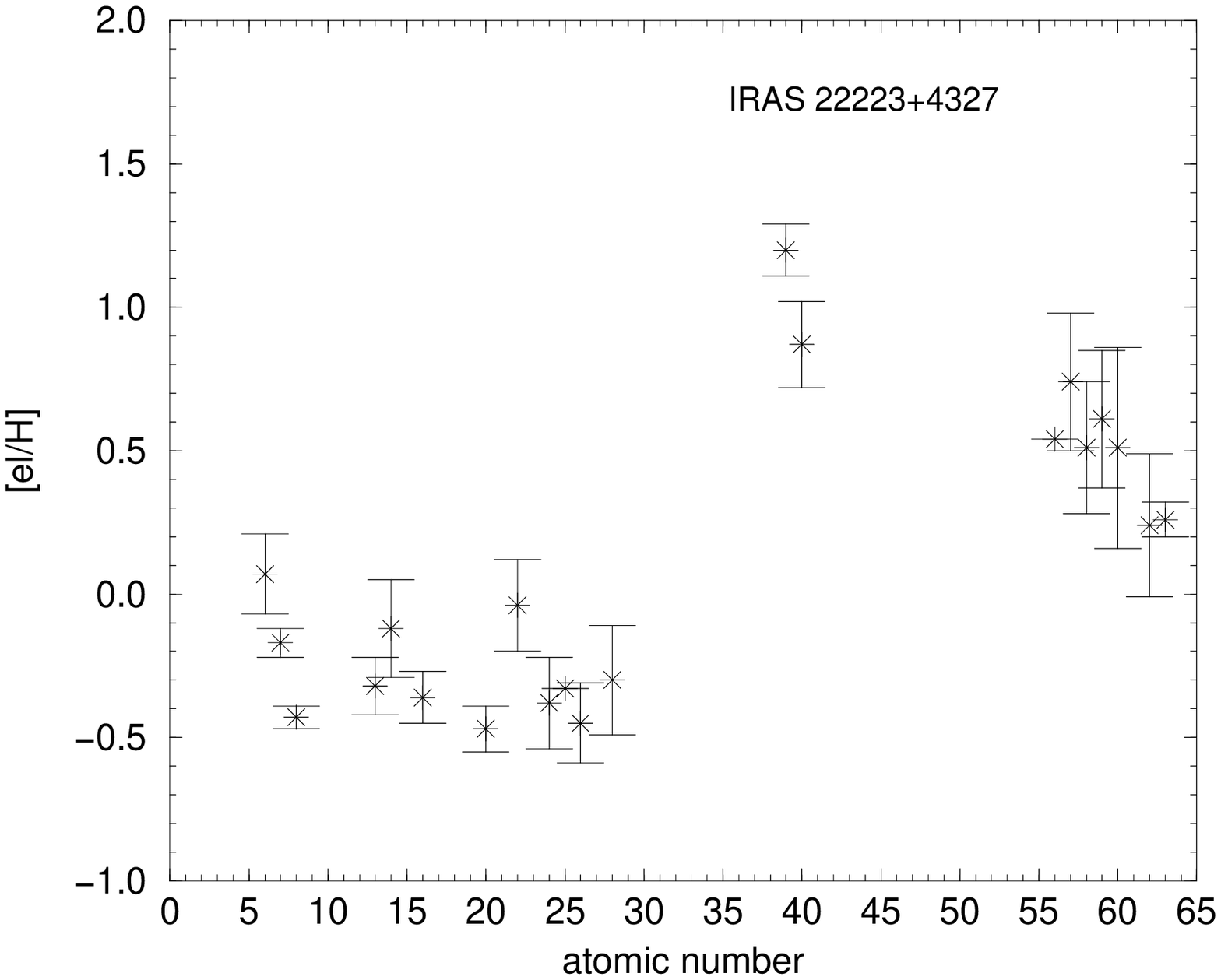}}
\mbox{\epsfxsize=0.48\textwidth 
\epsfbox{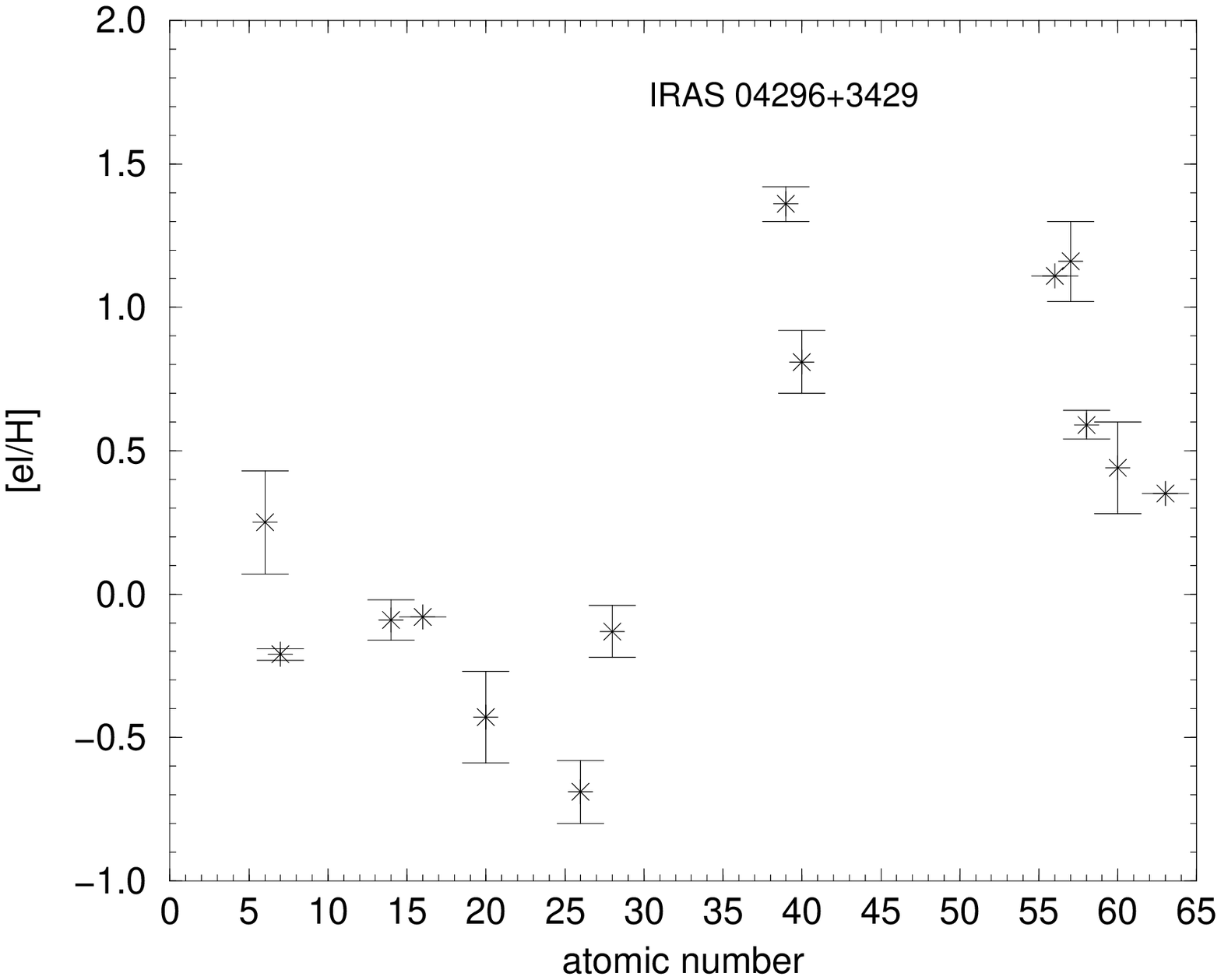}}
\caption{\label{iras2} The abundances of IRAS 22223+4327 and 
IRAS 04296+3429 relative to the solar values.}
\end{center}
\end{figure}

\subsubsection{Population}

Both \irt\ and \irv\ are iron deficient, [Fe/H]=$-$0.4 and $-$0.7
respectively.  Together with the value of their galactic latitude 
(b=$-$11.6 for
\irt\ and b=$-$9.1 for \irv\ ) this indicates that the 
IRAS-sources are low-mass objects of a
relatively old population. The radial velocity deduced from the spectra
are $V_{LSR} = -29 \pm 1.0 $\km  and $-62 \pm 1.0 $ \km
 for respectively
\irt\ and \irv\ which are similar to values found in the literature (Omont et
al.  1993). So far their is thus no evidence for radial velocity variations
and binarity of the programme stars.

\subsubsection{3rd dredge-up}


Both stars display a strong enrichment of carbon ([C/Fe] = +0.5 and +0.9), 
an indication that the third dredge-up was effective. 
During this third dredge-up material out of the
helium-burning shell is mixed into the stellar 
photosphere. Unfortunately, the O abundance is difficult to compute in this
temperature-gravity domain. The only lines available are the O triplet
at 6150 \ang, but these are heavily blended with a Fe~I and a Si~I line.
We only could determine the O abundance for the bright \irt\ and 
we obtain a C/O ratio of 1.3, but due to the large uncertainty of
the O abundance this ratio is not very accurate. From the atomic photospheric
lines alone the C/O ratio cannot be determined accurately enough to claim 
the objects to be real carbon stars (C/O $>$ 1).
The sum of the CNO nuclei is therefore also rather uncertain and compared to 
the expected value for an unevolved star of the same metallicity
we find only a $\Delta\sum$CNO of +0.05.

The most convincing argument for mixing products of the helium-burning
shell into the atmosphere are the large s-process elemental abundances.
For \irt\ and \irv\ we could derive abundances of 9 respectively 7
s-process elements and all turned out to be significantly overabundant, even
relative to the solar value. We will focus on the distribution of these elements
in a separate section.

For metal deficient stars the abundances of 
s-process-elements scale with Fe for $-$1.5$\le$[Fe/H]$\le$0.0 (see 
Wheeler 1989).  For \irt\ and \irv\ [s/Fe] is 1.0 and 1.4!
There is thus no doubt that both IRAS stars are post 3rd dredge-up
stars.

\subsection{Intermediate-mass elements (Al-Ni)}

For unevolved metal deficient stars a slight overabundance of 
the $\alpha$-elements reflects the chemical history of our Galaxy.
The solar neighbourhood displays
an increase of the [$\alpha$/Fe] ratio from 0.0 to +0.4 in the metallicity 
region
from about solar to [Fe/H]=$-$1.0 (Edvardsson et al. 1993)
When we take into account the error bars in the
$\alpha$-abundances, the overabundances of Si, S and Ni in \irv\ and Si and Ti
\irt\ do 
not seem to be abnormal. We may conclude that the 3rd dredge-up 
did not enhance significantly the $\alpha$ isotopes.

Also the odd elements are interesting tracers of internal nucleosynthesis
and structure. Theoretical models predict an enrichment of Al and especially
the Al/Mg ratio in
stars where Hot Bottom Burning (HBB) has taken place (Lattanzio et al.
1996). In the deepest layers of the convective 
envelope the temperatures may reach about 82 million K, and substantial
hydrogen burning via the CNO cycle will take place. 
Further, at these high temperatures $^{26}$Mg suffers substantial proton 
captures and produces $^{27}$Al during the interpulse phase. 
In theoretical evolutionary models, HBB is predicted only in
intermediate mass stars, since the temperature of the
bottom of the convective envelope only reaches in these models high enough
values for the synthesis to take place. Although the high 
C abundance already
indicates that HBB was not very effective in the IRAS objects, we carefully
analysed the
Al lines in our spectra of \irt. 
The strongest optical lines at 8773 \ang\ (multiplet number 9) are 
unfortunately heavily 
blended with the Phillips (2,0) band of the circumstellar C$_{2}$. We 
therefore based
or analysis of the lines of multiplet 10 (7835.3 and 7836.1 \ang) and 
multiplet 5 (6696.0 and
6698.7 \ang). The small overabundance of [Al/Fe] = $+$0.2 is not significant.
Also the $^{7}$Li resonance line at 6707 \AA\ is not detected, 
so we conclude that there is no evidence for even moderate HBB in \irt.
Unfortunately, there are no useful Mg lines in our spectra.

\subsection{s-proces element distribution}

Several theoretical studies of large nuclear reaction networks   
coupled with accurate AGB evolutionary codes exist
(Malaney 1987a,b; Busso et al. 1992,1995) that enable us to characterize 
the s-process in the IRAS sources based on the photospheric 
s-process element distribution.
The direct physical information on the efficiency of the internal 
nucleosythesis that can be deduced from the s-process 
distribution of an individual object, is unfortunately limited since the
predicted photospheric distribution is not only determined by the 
nucleosythesis 
itself, but also by the theoretically less understood dredge-up process 
(Busso et al. 1995). The parameters governing the outcome of such 
chemical evolutionary models do therefore not only consist of nucleosynthetic
quantities, but also of more ad-hoc adopted values governing the
stellar evolution like mass-loss, dredged-up mass, frequency of
dredge-ups etc. (Busso et al. 1992).

The  ratio of light (Sr, Y, Zr) to heavy 
(Ba, La, Nd and Sm) s-elements gives a measure of the neutron 
exposure rate which is defined as $\int_{0}^{t} N_{n}(t') V(t') dt' $
where $N_{n}$ is the neutron density, $V$ the relative velocity of the
neutrons and the seed nuclei. The pulsed neutron irradiation during
the AGB evolution is parameterized by defining a mean neutron exposure
$\tau_{0}$ defined a $\tau_{o} = \Delta\tau\/(-\ln r)$ with $\Delta\tau$
the neutron exposure rate of a particular pulse and r the overlap factor
which is the fraction of the inter-shell material that remains in the 
neutron-exposed region (Ulrich 1973).
In most calculations, the neutron density and the overlap factor are 
adopted as constants and the efficiency of s-processing is indicated by one 
parameter $\tau_{0}$.

For  \irt\ we obtained the most complete s-process distribution. In the 
temperature-gravity domain of the moderately deficient programme stars, 
the Sr abundance is extremely difficult to measure due to the lack 
of weak lines. We therefore did not take this element into account to
determine the [ls/Fe] ratio but following Busso et al. (1995) no correction 
factor is needed to account for unobserved elements from the light
s-process trio.
The mean abundance of the light elements Y and Zr 
is [ls/Fe] = +1.5 while for the heavy elements Ba, La, Nd and 
Sm [hs/Fe] = +0.9, hence the [hs/ls] = $-$0.6.
Following Fig. 6 of Busso et al. (1995) the mean neutron exposure
of the object can be estimated to be in between 0.2 and 0.25 $\rm{mbarn^{-1}}$.

\irt\ is located in this diagram in the locus of the Carbon stars
and the high value of [ls/Fe] indicates a high mixing ratio between
dredge-up material and residual mass of the envelope (Busso et al. 1995).
This is not surprising since the post-AGB star is Carbon rich.

More surprising is, however, the low value of the mean neutron exposure
of the object given its low metallicity. Indeed, several observational
evidences exist that the neutron exposure increases with lower
metallicity. In Fig. 1 of Busso et al. (1995), where they show
the measured [hs/ls] values as a function of the iron abundance [Fe/H]
of a sample of intrinsic and extrinsic S stars and Ba stars, a clear 
correlation is seen with an increase of 0.2 dex in [hs/ls] for a drop 
in metallicity from 0 to $-$0.5. While the mean trend of the Ba-stars 
indicate that an object with metallicity
of $-$0.5, should have a [hs/ls] of $+$0.2, this ratio is only $-$0.6 for
\irt!
Since the metallicity of Carbon stars is difficult to measure, 
no observational material is present for comparison.

\begin{figure}
\begin{center}
\mbox{\epsfxsize=0.48\textwidth
\epsfbox{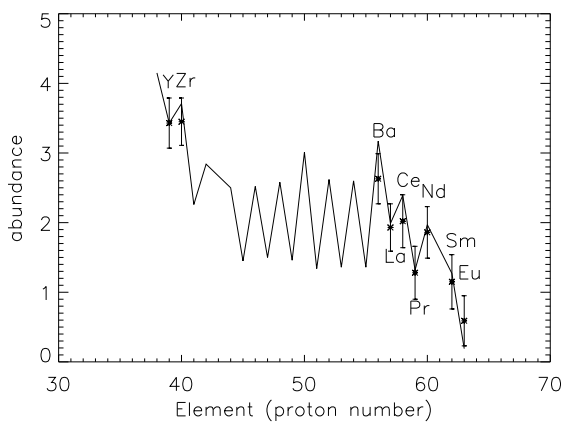}}
\caption{\label{sprocess} Comparison between the observed s-proces
abundance distribution of \irt\ and model predictions from Malaney (1987b)
for an exponential neutron exposure characterized with $\tau_{0} = 
0.2 \rm{mbarn^{-1}}$ and
a neutron density of $N_{n} = 10^{8}$ $\rm{cm^{-3}}$.}
\end{center}
\end{figure}

In order to characterize more precisely the s-process efficiency, we
made use of the abundance tables for s-process nucleosynthesis of
Malaney (1987b). The tables list element abundance enhancements 
for exponential distribution of neutron exposures for different
values  $\tau_{0}$ using fixed neutron densities of $10^{8}$ and $10^{12}$
$\rm{cm^{-3}}$. To quantify the comparison between the observations
and models we used the goodness-of-fit procedure as defined by Cowley
\& Downs (1980) and often used in the literature. For a description of
the method we refer to Smith et al. (1997). In the comparison, we used
the abundances of all the measured species : Y (Z=39), Zr(40), Ba(56), La(57),
Ce(58), Pr(59), Nd(60), Sm(62) and Eu(63). For the uncertainty on the measured
abundances we added quadratically a fixed value of 0.3 to the internal
accuracy given in Table~5, this to account for
uncertainties on the model atmosphere and uncertain $\log gf$ values.
Moreover, for species with less than 5
useful lines measured, we adopt an internal accuracy of 0.2 dex. We corrected
the observed number densities for the initial abundance by adopting
[s-process/Fe] = 0 and a metallicity of -0.5, but since the overabundance
is large, this correction only marginally influences the result. For Eu, which
is primarily an r-process element with a different chemical history
than the s-process elements, we used the results of Woolf et al. (1995)
to estimate the initial abundance. 
In this method, the quality of the fit is given by the quantity $S^{2}$,
with a lower value expressing a better fit. In Fig. \ref{sprocess}
we plot our corrected abundance distribution together with Malaney's (1987b)
model predictions for an exponential neutron exposure with $\tau_{0}$
of $0.2\ \rm{mbarn^{-1}}$ ($N_{n} = 10^{8} \rm{cm^{-3}}$) with an $S^{2}$ value 
of 0.6. 
For higher neutron efficiencies, the $S^{2}$ parameter increases fast 
(e.g. for $\tau_{0}$ = 0.3, $S^{2}$ is already more than tripled) and for
the $\tau_{0}$ = 0.1 model we have $S^{2}$ = 3.7. Also the model with
$\tau_{0}$ = 0.05 gives a good fit but the absolute enhancement
of the s-process elements is, in this model, predicted to be too low in
comparison with the observed values.
We can conclude that the distribution of s-process elements in \irt\ points 
to a low s-process efficiency characterized by 
a $\tau_{0} \leq 0.2\ \rm{mbarn^{-1}}$.

The s-process distribution of \irv\ also points to a low s-process
efficiency with the best fit again obtained for the models with $\tau_{0}
\leq 0.2\ \rm{mbarn^{-1}}$. In this object, only 7 elements were taken
into account in the fitting.

\section{Discussion}

\begin{table*}[t!]
\caption{\label{micron}Summary of spectral features in the best studied
 {\bf 21 \mic\ sources}, emphasizing correlation with carbon.}
\begin{center}
\scriptsize
\begin{tabular}{|l||c|c|c|c|c|c|c|c|c|c|r|} \hline
\rule[-3mm]{0mm}{8mm}name & other names & b & V & 21 \mic & C$_2$ & 3.3
\mic & 3.4 \mic & 30 \mic & molecular lines & $T_{\rm eff}$ & references \\
\hline

\rule[-3mm]{0mm}{8mm}
IRAS~04296+3429 & & $-$9.1 & 14.2 & Y & Y  & Y & Y &     & CO, HCN, CN &
7000K & 1, 3, 5, 8, 12 \\ 
\rule[-3mm]{0mm}{8mm}
IRAS~05113+1347 & & $-$14.3 & 12.4 & Y & Y  & Y & N &     & CN & & 1, 7, 
8, 11 \\
\rule[-3mm]{0mm}{8mm}
IRAS~05341+0852 & & $-$12.2 & 12.8 & Y & Y  & Y & Y &     & CN & 6500K & 11,
14 \\ 
\rule[-0mm]{0mm}{5mm}
 & HD~56126 & & & &  & & & & & & \\
\rule[-3mm]{0mm}{3mm}
{\raisebox{1.5ex}[0pt]{IRAS~07134+1005}} & SAO~96709 &
{\raisebox{1.5ex}[0pt]{+10.0}} & {\raisebox{1.5ex}[0pt]{8.2}} &
{\raisebox{1.5ex}[0pt]{Y}}  & {\raisebox{1.5ex}[0pt]{Y}}   
& {\raisebox{1.5ex}[0pt]{Y}} & {\raisebox{1.5ex}[0pt]{N}} &
{\raisebox{1.5ex}[0pt]{?}}  & {\raisebox{1.5ex}[0pt]{ CO, HCN, CN}} &
{\raisebox{1.5ex}[0pt]{7000K}}  & {\raisebox{1.5ex}[0pt]{2, 3, 4, 10, 17}} \\
\rule[-0mm]{0mm}{5mm}
 & HD~187885 & & &  & & & & & & 7500K & \\
\rule[-3mm]{0mm}{3mm}{\raisebox{1.5ex}[0pt]{IRAS~19500$-$1709}} & SAO~163075 &
{\raisebox{1.5pt}[0pt]{$-$21}} & {\raisebox{1.5ex}[0pt]{9.2}} &
{\raisebox{1.5ex}[0pt]{weak}} & {\raisebox{1.5ex}[0pt]{N}}  &
{\raisebox{1.5ex}[0pt]{   }} &
{\raisebox{1.5ex}[0pt]{   }}  & {\raisebox{1.5ex}[0pt]{   }} &
{\raisebox{1.5ex}[0pt]{No CN}}  & 8000K &{\raisebox{1.5ex}[0pt]{ 9, 15}} \\  
\rule[-3mm]{0mm}{8mm}
IRAS~20000+3239 &  & +1.2 & 13.3 & Y & Y &     &     & Y & CO, HCN, CN & &
1, 7, 8, 11, 12 \\
\rule[-3mm]{0mm}{8mm}
 & AFGL2688 & $-$6.5 & 14.0 & Y & Y &      &     & Y & HCN, CN & &  13 \\
\rule[-3mm]{0mm}{8mm}
IRAS~22223+4327 & DO~41288 & $-$11.6 & 9.7 & Y & Y  &     &     &     & CO,
HCN, CN & 6500K & 1, 7, 8, 11, 12 \\
\rule[-0mm]{0mm}{5mm} 
 & HD~235858 & & & & & & &  & & & \\
\rule[-3mm]{0mm}{3mm}
{\raisebox{1.5ex}[0pt]{IRAS~22272+5435}} & SAO~34504 &
{\raisebox{1.5ex}[0pt]{$-$2.5}} & {\raisebox{1.5ex}[0pt]{8.7-9.5}} &
{\raisebox{1.5ex}[0pt]{Y}} &  {\raisebox{1.5ex}[0pt]{Y}} & 
{\raisebox{1.5ex}[0pt]{Y}} &
{\raisebox{1.5ex}[0pt]{Y}} &  {\raisebox{1.5ex}[0pt]{Y}} &
{\raisebox{1.5ex}[0pt]{CO, HCN, CS, CN}} &  &  {\raisebox{1.5ex}[0pt]{1, 3, 6,
8, 17}} \\ 
\rule[-3mm]{0mm}{8mm}
IRAS~22574+6609 & & +6.0 & & Y &      &     &     &     &      & & 5
\\ 
\rule[-3mm]{0mm}{8mm}
IRAS~23304+6147 &  & +0.6 & 1.31 & Y & Y &     &     & Y & CO, HCN, CN & &
1, 3, 5, 8, 16 \\
\hline
\end{tabular}
\end{center}
\footnotesize{
1. Bakker et al. 1995 
2. Bakker et al. 1997
3. Hrivnak et al. 1989a 
4. Hrivnak et al. 1989b
5. Hrivnak \& Kwok 1991a
6. Hrivnak \& Kwok 1991b
7. Hrivnak et al. 1994
8. Hrivnak 1995
9. Justtanont et al. 1996
10. Klochkova 1995
11. Kwok et al. 1995
12. Omont et al. 1993
13. Omont et al. 1995
14. Reddy et al. 1997
15. Van Winckel et al. 1996a
16. Woodsworth et al. 1990
17. Zuckerman et al. 1986}
\end{table*}
\normalsize

The chemical composition indicates
\irt\ and \irv\ to be 'bona-fide' post-AGB stars: they display an enrichment 
of carbon with C/O  higher than solar and especially high overabundances
of s-process elements. For both stars the predicted heavy element abundances
for an exponential distribution of neutron exposures with low value of the mean
neutron exposure $\tau_{0} = 0.2\ \rm{mbarn^{-1}}$ fit the
observed s-process abundances best.

Note that not all stars, thought to be in-between the AGB and PN
stage of stellar evolution, have all these chemical properties. E.g. HD~133656
([Fe/H]=$-$1) shows CNO enrichment, but no overabundance of the 
s-process-elements (Van Winckel et al. 1996b).
Also SAO~239853 ([Fe/H]=$-$0.8---$-$1.0) (Van Winckel 1997) and 
HR~4912 (Luck \& Bond 1989) are thought to be post-AGB stars without
an enrichment of the s-process-elements. 
On the contrary, other 21 \mic\ stars like IRAS~07134+1005
([Fe/H])=$-$1.17; Klochkova 1995) and IRAS~05341+0852 
([Fe/H] = $-$0.9; Reddy et al. 1997) do also show overabundances 
of s-process-elements.

The two IRAS sources \irt\ and \irv\  belong to the small 
group of stars displaying the unidentified 21 \mic\ emission
feature.  The other best studied 21 \mic\ sources
are listed in Table~\ref{micron}
(Kwok et al. 1989; Hrivnak \& Kwok 1991a,b; Kwok 1993; Hrivnak 1995;
Justtanont et al. 1996). Some of the sources are resolved in the 8-13 \mic\
window (Meixner et al. 1997)
Note that the table does not include the 
new possible candidates given by Henning et al. (1996).

Table~\ref{micron}  shows that the 21 micron 
sources have some common properties:\\
1. they are post-AGB stars with a central star of spectral type F or G and
luminosity class I (Kwok et al. 1995)\\
2. all have double-peaked energy distributions (Hrivnak 1995)
\\
3. all are carbon-rich objects based on their optical, infrared or millimetre
spectra \\
4. all seven of which have been observed show  HCN molecular lines\\ 
5. five sources display the 3.3 \mic\ emission band and three of them also an
unusually strong 3.4 \mic\ emission band\\
6. four display the 30 \mic\ emission band (Hrivnak 1995; Omont et al. 1995)\\
7. all the 21 micron sources of which the optical spectra
have been analysed (\irt, \irv, IRAS~07134+1005, IRAS~05341+0852 and HD~187885)
show an overabundance of s-process-elements, but also the reverse is true : all
field post-AGB stars known to be s-process enriched, also 
display the 21 \mic\ feature.

The carbon-rich chemistry, as evidenced by the C$_2$ circumstellar molecular
lines and the photospheric analyses corroborates the suggestion of 
Hrivnak et al. (1989a) that carbon
is a major constituent of the molecule producing the 21 \mic\ emission feature.
So far the only
star that evolved hot enough to display photospheric helium lines is HD~187885
(Van Winckel et al. 1996a) which only has a very weak 21 \mic\ feature. The
effective temperature of 8000 K marks probably the region where the carriers
of the feature are destroyed.
High-resolution spectroscopy of more central stars and certainly accurate
determinations of the effectieve temperatures are needed to
test several theories concerning the nature of 
the IR features and the influence of changing properties of the stellar
radiation field on them (Buss et al. 1990; Begemann et al. 1997).

So far, the study of heavy element nucleosynthesis to constrain 
3rd drege-up models, concentrated on two groups of stars :
the AGB stars (mainly Carbon stars, but also M stars that display s-process
enhancements) and the Ba stargroup (Ba giants, Tc poor S stars, 
CH stars etc.). 
In the latter group the s-process enhancement is generally accepted to originate
from mass-transfer episodes. These objects are binaries with one 
component being a cool old white dwarf (WD). 
The chemical peculiarities were build up when the star, which is now 
the primary,
accreted s-process element enhanced material from the companion, 
which then was an AGB star and now is a cool white dwarf.
The s-process distribution seen in these stars is thus not a direct result
of internal mixing of the Ba-star itself, but a result of a not well
understood accretion process of s-process enhanced material. 

The study of the s-process element distribution of AGB stars themselves 
is made extremely difficult by the large photospheric molecular opacity that 
makes the modelling difficult and the detection of several species 
impossible (e.g. Utsumi 1985). 

A systematic study of post-AGB stars in general and of the 21 \mic\
post-AGB stars in particular can in this debate contribute 
invaluable information. Indeed, not only the temperature-gravity domain 
of these stars makes it possible to
detect a large number of s-elements by their atomic lines, but also the
photospheric molecular opacity is negligible. Since the photospheric abundance
is a direct result of the dredge-up process (contrary to the case of the
Ba-stars) and a fair number of 21 \mic\ stars are now known, 
a systematic study of the s-process distributions can be used directly 
to constrain evolutionary models. In this first study
the s-process distribution is characerized for \irt\ and \irv\ both 
with mean neutron exposure  $\tau_{0} \leq 0.2\ \rm{mbarn^{-1}}$  
and for another objest IRAS~05341+0852, Reddy et al. (1997) find the 
$\tau_{0} \simeq 0.3\ \rm{mbarn^{-1}}$. This value
is certainly much lower than what might be expected from their metallicity
if the 21 \mic\ stars would follow the same metallicity-neutron exposure
relation of the Ba-stars (see Fig. 1 Busso et al. 1995).
It is clear that a more homogeneous study of more sources is 
needed to compare carefully the different groups of stars.

\begin{acknowledgements}

LD and HVW acknowledge support from the Science Foundation of Flanders.
EJB acknowledges support form the National Science Foundation (Grant 
No. AST-9315124). This research has made use of the SIMBAD database, 
operated at CDS, Strasbourg, France and the Vienna Atomic Line Data-base
(VALD), Vienna, Austria. The authors thank the anonymous referee for his
interesting comments.

\end{acknowledgements}

\end{document}